\documentclass[epj]{svjour}
\usepackage{graphicx}
\usepackage{amsmath}
\usepackage{times}
\usepackage{amssymb}
\usepackage{bm}%
\usepackage{hyperref}%
\usepackage{cite}
\usepackage{color}
\begin{document}
\title{The kinetic fragility of liquids as manifestation of the elastic softening}
\author{F.\@ Puosi \inst{1} \thanks{Present 
 address: Laboratoire de Physique de l'\'Ecole Normale Sup\'erieure de Lyon,  UMR CNRS 5672, 46 all\'ee d'Italie, 69007 Lyon, France.} \and D.\@ Leporini \inst{1} \inst{2}
\thanks{E-mail: \email{dino.leporini@df.unipi.it}}%
}                     
%
%
\institute{Dipartimento di Fisica ``Enrico Fermi'', 
Universit\`a di Pisa, Largo B.\@Pontecorvo 3, I-56127 Pisa, Italy \and IPCF-CNR, UOS Pisa, Italy}
\date{Received: date / Revised version: date}
%
\abstract{
We show that the fragility { $m$}, the steepness of the viscosity and relaxation time close to the vitrification, increases with the degree of elastic softening{, i.e. the decrease of the elastic modulus with increasing temperature,} in universal way. 
{ This provides a novel connection between the thermodynamics, via the modulus, and the kinetics.} 
The finding is evidenced by numerical simulations and comparison with the experimental data of glassformers with widely different fragilities ($33 \le m \le 115$), leading to a fragility-independent  elastic master curve extending over eighteen decades in viscosity and relaxation time. The master curve is accounted for by a cavity model { pointing out the roles of both the available free volume and the cage softness}.   
A major implication of our findings is that ultraslow relaxations, hardly characterised experimentally, become predictable by linear elasticity. As an example, the viscosity of supercooled silica is derived over about fifteen decades with no adjustable parameters. 
} 
\maketitle
\section{Introduction}
\label{intro}
Glassformers are classified in terms of their kinetic fragility, as quantified by the fragility index $m \equiv \partial \log \tau_\alpha/ \partial (T_g/T) |_{T_g/T=1}$, where $\tau_\alpha$ and $T_g$ denote the structural relaxation time and the glass transition temperature, respectively \cite{AngellJNCS91}. 
Fragility is a measure of the degree of departure from the Arrhenius scaled temperature dependence, which is weak for "strong" glassformers and quite apparent for "fragile" ones \cite{StilliDebe01,EdigerHarrowellJCP12}. It is worth noting that the terminology "strong" and "fragile" was introduced in relation to the evolution of the short-range order close to $T_g$ \cite{BerthierBiroliRMP11}. {  Different, often controversial, viewpoints concerning the link of fragility with structure, thermodynamics and dynamics have been reported\cite{AngelNgai00,TarjusJPCM05,DudowiczEtAl08,NgaiBook,BerthierBiroliRMP11,SokolovFragility13,RoyallWilliamsPhysRep15,EdigerHarrowellJCP12,McKennaFragilityThermodynJCP01,FragilityMauroSciRep15}}.

Here, we argue that  the fragility is related to the mechanical properties of the {\it liquid}, and
the structural relaxation time $\tau_\alpha$ (or viscosity $\eta$) is an universal function of the linear elastic modulus $G_p$, irrespective of the kinetic fragility. Then, different  fragilities just reflect different
{ degrees of elastic softening, i.e. the decrease of the elastic modulus with increasing temperature, being} weak for strong glassformers and more marked for fragile ones. This provides a connection between the thermodynamics, via $G_p$, and the kinetics. 

{ The present paper contributes to the living 
discussion on the role of elasticity and internal stresses in the structural 
relaxation of supercooled liquids.}
It has been proposed that structural relaxation in deeply supercooled liquids proceeds via the accumulation of Eshelby events, i.e. local rearrangements that create long-ranged and anisotropic stresses in the surrounding medium \cite{LemaitrePRL14}.
Fragility and elastic softening have been correlated \cite{GranatoFragilityElasticityJNCS02} in the framework of the interstitialcy \cite{GranatoPRL92} and the conventional elastic \cite{DyreRevModPhys06,Nemilov06} models of the glass transition.  Theoretical work supports  the conclusion that glass elasticity affects the fragility of supercooled liquids \cite{WyartPNAS2013}, see also \cite{NovikovEtAl05,Novikov04}. { An elastically collective nonlinear Langevin equation has been derived and successfully compared to van der Waals liquids  by mapping real molecules to an effective hard sphere fluid \cite{SchweizerJPCLett13,SchweizerElastic1JCP14,SchweizerElastic2JCP14}. The approach has  diminished quantitative accuracy when the fragility decreases. Recently, quantitative relations between cooperative
motion, elasticity, and free volume  have been found in model polymeric glass-formers \cite{DouglasStarrPNAS2015}}.

Universal aspects of the caging effects in viscous liquids are central to the present results. Close to the glass transition, particles tend to be trapped in transient cages formed by their nearest neighbours with subsequent slowing down of their mean square displacement $\langle r^2(t)\rangle$ within time $t$ \cite{BerthierBiroliRMP11}. The particles rattle about in the cage on picosecond time scales with mean square amplitude $\langle u^2\rangle$ and are
later released with average escape time $\tau_\alpha$.  Structural relaxation and cage rattling are correlated and one finds the universal master curve \cite{wolyNatPhys}:
\begin{equation}
\log \tau_\alpha = {\alpha} + {\beta} \, \frac{1}{\langle u^2\rangle} + {\gamma} \, \frac{1}{\langle u^2\rangle^{2}}
\label{parabolaMD}
\end{equation}
${\alpha}$, ${\beta}$ and ${\gamma}$ are suitable constants independent of the kinetic fragility. Eq.\ref{parabolaMD} has been tested on experimental data \cite{wolyNatPhys,OttochianLepoJNCS11,UnivPhilMag11,CommentSoftMat13,SokolovNovikovPRL13} and numerical models of polymers \cite{wolyNatPhys,lepoJCP09,Puosi11,Puosi12SE}, colloids \cite{DeMicheleDelGadoLepo11} and atomic liquids \cite{lepoJCP09,PuosiSpecialIssueJCP13}. The fast rattling motion of particles during the trapping periods in {liquids} has strong analogies with the oscillatory  {\it elastic} behaviour  of particles in crystalline and amorphous solids, a major difference being that liquids exhibit {\it transient} elasticity terminated by the structural relaxation \cite{Puosi12}.

{ In this paper} we first characterize polymer melts with {\it different}  fragilities by extensive Molecular-Dynamics (MD) simulations. We evidence that, irrespective of the fragility, the relaxation time exhibits the {\it same} scaling with the elastic modulus:
\begin{equation}
\log  \tau_\alpha  =  \Upsilon_0 + \Upsilon_1 \left ( \frac{G_p }{T} \right ) + \Upsilon_2  \left (\frac{G_p }{T} \right )^2
\label{masterG}
\end{equation}
where $\Upsilon_{0}$, $\Upsilon_1$ and $\Upsilon_2$ are constants {\it independent} of the fragility. We show that the MD results do not support the assumption that the kinetic unit is embedded in an elastic {\it continuum} (EC) and develop a novel elastic model combining packing effects with elasticity. 
Furthermore, we reveal the elastic scaling in glassformers with intermediate and high fragilities ($33 \le m \le 115$) and collapse the experimental relaxation times (or viscosity) over about eighteen decades on a universal master curve given by Eq. \ref{masterG} recast in terms of the reduced quantity: 
\begin{equation}
X = \frac{G_p  T_g}{G_{p g} T}
\label{reducedX}
\end{equation}
where $G_{p g} \equiv G_p(T_g)$. Finally, to test the robustness of the scaling, we predict the viscosity of the strong glassformer SiO${}_2$ ($m=20$) by its linear elasticity with {\it no} adjustable parameters. We find excellent agreement over a range spanning about fifteen orders of magnitude where the viscosity exhibit deviations from the Arrhenius behaviour. 

We compare our findings to the conventional elastic models of the glass transition \cite{DyreRevModPhys06,Nemilov06}. Their main result is:
\begin{equation}
\log  \frac{\tau_\alpha}{ \tau_{\alpha 0}} =  G_p V^\star / k_B T
\label{ElasticModel} 
\end{equation}
where $\tau_{\alpha 0}$ and $V^\star$ are {\it adjustable} parameters. With respect to these models, we provide totally new insight (the fragility- independent scaling), fix known problems with fragile liquids \cite{Rouxel11,NelsonJCP09} and improve the agreement with the paradigmatic strong liquid SiO${}_2$ without {\it any} adjustable parameter. 

\begin{figure}[t]
\begin{center}
\includegraphics[width=8cm]{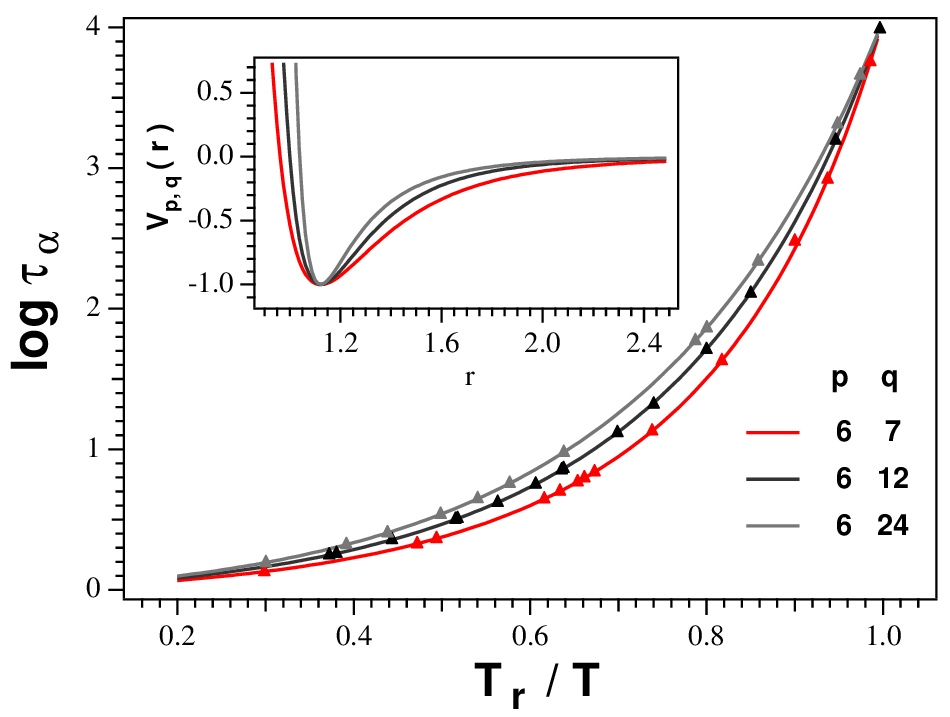}
\end{center}
\caption{ Reduced temperature dependence of the relaxation time of trimers ($M=3$) with different forms of the interacting potential between non-bonded monomers (insert). $T_r$ is the temperature where $\tau_\alpha = 10^4$. The number density is $\rho = 1.033$.
}
\label{fig00}
\end{figure}

\section{MD simulations}
\label{SimMD}
We perform extensive molecular-dynamics (MD) simulations of a melt of fully-flexible linear chains of $M$ soft spheres (monomers,  $N \simeq 2000$ in total).  The interacting potential between non-bonded monomers has the form
\begin{equation}
V_{p,q}(r)= \frac{\epsilon}{(q-p)} \left [ p \left (\frac{\sigma^\star}{r} \right )^q - q \left (\frac{\sigma^\star}{r} \right)^p \, \right ]
\label{vpot}
\end{equation}
with $\sigma^\star = 2^{1/6} \sigma$. Changing the $p$ and $q$ parameters does not affect the position $r = \sigma^\star $ and the depth $\epsilon$ of the potential minimum but only the steepness of the repulsive and the attractive wings (see Fig.\ref{fig00} and Supplementary Information (SI)). The potential $V_{p,q}(r)$ has adjustable anharmonicity which, according to studies on atomic liquids \cite{BordatNgai04}, is able to tune the kinetic fragility. Fig. \ref{fig00} shows that this occurs for the polymer melt too.
All quantities are in reduced units (Boltzmann constant $k_B=1$): length in units of $\sigma$, temperature in units of $\epsilon/ k_B$, and time in units of $\sigma \sqrt{m/\epsilon}$, where $m$ is the monomer mass. The potential is cut and shifted to zero by $U_{cut}$ at $r=2.5$. The bond length is $b=0.97$. 
For each form of the potential $V_{p,q}(r)$ several physical states are collected by changing the temperature $T$, the number density $\rho$ and the number of monomers per chain $M$. Further details about the MD simulation are given in SI where all the states characterised by their elasticity ($\sim 100$) are also listed. 

 The collective elastic dynamics is described by the transient elastic modulus of a volume $V$, $G(t)$, being expressed by the correlation function \cite{Puosi12}:
\begin{equation}
\label{gt}
G(t) = \frac{V}{k_B T} \langle \sigma_{xy}(t_0) \sigma_{xy} (t_0+t) \rangle. 
\end{equation}
$\sigma_{xy}$ is the off-diagonal component of the stress tensor:
\begin{equation}
\label{stresstensor}
\sigma_{xy} = \frac{1}{V} \Bigg (    \sum_{i=1}^N \Bigg [  m v_{xi} v_{yi} + \frac{1}{2} \sum_{j \ne i} r_{xij} F_{y ij} \Bigg ] \Bigg )
\end{equation}
where $v_{\alpha k}$,  $F_{\alpha  k l}$, $r_{\alpha  k l}$  are the $\alpha$ components of the velocity of the $k$th monomer with mass $m$, the force between the $k$th and the $l$th monomer and their  separation, respectively. The symbol $\langle \cdots \rangle$ represents the canonical average.
The monomer mean square displacement is defined as: $\langle r^2(t)\rangle = N^{-1}  \langle \sum_{j=1}^{N} [\bm{r}_j(t) - \bm{r}_j(0)]^2  \rangle$, where  $\bm{r}_j(t)$ is the position of the $j$-th monomer at time $t$, the sum runs over the total number of $N$ monomers. 
At $t^\star \simeq 1.023$ early detrapping of the monomers from their cages occurs and the quantity $\partial \log\langle r^2(t)\rangle/ \partial \log t$ shows a well-defined minimum
($t^\star$ is independent of the physical state in the present model) \cite{wolyNatPhys,lepoJCP09,Puosi11,Puosi12SE,PuosiSpecialIssueJCP13}.  We define the mean square amplitude of the position fluctuations of the monomers in the cage as  \cite{wolyNatPhys,lepoJCP09,Puosi11,Puosi12SE,PuosiSpecialIssueJCP13,DeMicheleDelGadoLepo11}: 
\begin{equation}
\langle u^2\rangle \equiv \langle r^2(t^\star)\rangle
\label{STMSD}
\end{equation}
One finds \cite{Puosi12}
that in a time $t^\star$ {\it mechanical equilibration} is reached, the total force on each particle vanishes, and the off-diagonal stress correlation function $G(t)$ has reached the intermediate-time plateau 
setting the linear shear modulus:
\begin{equation}
G_p \equiv G(t^\star)
\label{Gp}
\end{equation}
The incoherent intermediate scattering function $F_s(q_{max},t)$ is defined as $F_s(q, t) =  N^{-1}  \langle \sum_{j=1}^{N}  exp\{-i \bm{q}\cdot [\bm{r}_j(t) - \bm{r}_j(0)] \}  \rangle $,  $q_{max}$ being the q-vector of the maximum of the static structure factor 
\cite{LariniCrystJPCM05,PrevostoEtAl04,AndreozziEtAl98,AndreozziEtAl99}.
The structural relaxation time $\tau_\alpha$ is defined by the relation $F_s(q_{max},\tau_\alpha)= 1/e$.  

\begin{figure}[t]
\begin{center}
\includegraphics[width=8cm]{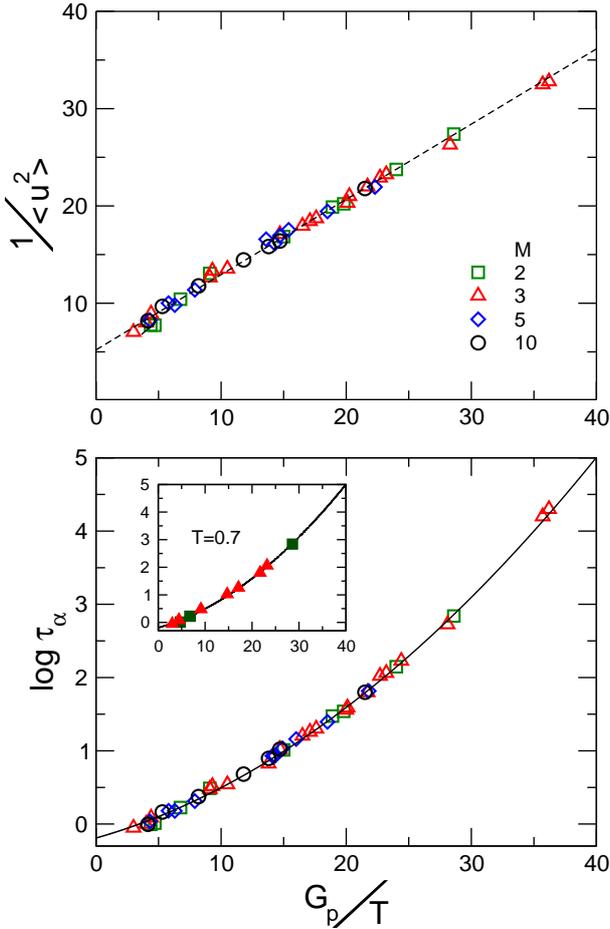}
\end{center}
\caption{Correlation between the mean square rattling amplitude of the monomer in the cage $\langle u^2\rangle$ (top) and the relaxation time $\tau_\alpha$ (bottom) with the ratio $G_p/T$ from MD simulations. The dashed line in the top panel is  Eq.\ref{rettaelastica} with best-fit parameters $\ell= 0.77(2) $ and $\zeta^2 = 0.192(4)$.
The solid line in the bottom panel is Eq.\ref{masterG} with $\Upsilon_0 = -0.191(8)$, $\Upsilon_1 =0.048(3)$,  $\Upsilon_2 =0.0020(1)$, as obtained by the combination of Eq.\ref{parabolaMD} (${\alpha} =-0.424(1), {\beta} = 2.7(1) \cdot 10^{-2}, {\gamma} =  3.41(3) \cdot 10^{-3}$ \cite{wolyNatPhys}) with Eq.\ref{rettaelastica} . No adjustable parameters are allowed. The insert shows that the scaling holds for isothermal data as well (the solid line is the same of the main panel).}
\label{fig1}
\end{figure}

\section{Results}
\subsection{Elastic scaling in simulations of a polymer melt.}

First, we investigate the relation between the mean square amplitude of the cage rattling $ \langle u^2\rangle$ and the elasticity in polymer systems with {\it different} fragility. It is known that $ \langle u^2\rangle$ does sense the fragility \cite{Ngai04}. Fig.\ref{fig1}(top) summarises the results and evidences a {\it fragility-independent} master curve:
\begin{equation}
 \frac{1}{\langle u^2\rangle} =  \frac{1}{\zeta^2} + \frac{G_p \; \ell}{k_B T}
\label{rettaelastica}
\end{equation}
The two length scales $\ell$ and $\zeta$ are nearly constant, most probably due to the limited changes of the local structures in virtue of the high packing of the investigated states \cite{HallWoly87,StarrEtAl02,BerniniJCP13,VoroBinarieJCP15,BarbieriGoriniPRE04}.  This means that $\ell$ and $\zeta$ depend on the density and the interacting potential - both affecting the fragility \cite{BordatNgai04,Sastry01} - and the temperature in much weaker way than the elastic modulus. This suggests that the fragility-dependence of $ \langle u^2\rangle$ occurs mainly via the elasticity. 

Fig.\ref{fig1} (bottom) shows that the elastic scaling also collapses the relaxation time of polymer melts with different fragilities, e.g. see Fig.\ref{fig00}, on a {\it fragility-independent} master curve. The insert shows that the scaling holds also under isothermal conditions and exposes the wide range of elastic moduli under consideration.
The master curve in Fig.\ref{fig1}(bottom) has the form of Eq.\ref{masterG} and is achieved by combining the best-fit of Eq.\ref{rettaelastica} with Eq.\ref{parabolaMD} {\it without adjustment}.  Note that the master curve is not a straight line, namely it differs from the prediction of the conventional elastic models, Eq.\ref{ElasticModel}, confirming - as reported \cite{Rouxel11,NelsonJCP09} - that they face problems when dealing with fragile glassformers like the present simulated ones, see Fig.\ref{fig00}.

Eq.\ref{rettaelastica}, cannot be rationalised within the picture of a particle embedded in an elastic {\it continuum} (EC). In fact, a particle embedded in EC with shear modulus $G_p$  undergoes position fluctuations with mean square amplitude $\langle u^2_\mathrm{EC}\rangle$  given by \cite{DyreRevModPhys06,SchmidtEtAl97,VanZantenRufener00,Maradudin_DebyeModel,Dyre04}:
\begin{equation}
\langle u^2_\mathrm{EC}\rangle = \frac{k_B T}{G_p \, \ell_\mathrm{EC}} 
\label{FDT_Elastico}
\end{equation}
where $\ell_\mathrm{EC}$ is comparable with the particle size. Fig.\ref{fig1}(top) shows that Eq.\ref{FDT_Elastico} is inadequate if applied to a particle trapped in a discrete environment. The disagreement is anticipated since Eq.\ref{FDT_Elastico} relies on the affinity of the microscopic and the macroscopic displacements, a feature which breaks down in discrete systems \cite{DePablo04,Maloney06,BarratPRE09,Wallace72}.
 We present in Sec.\ref{elasticcavity} a theoretical treatment which includes the microscopic discreteness of the system and correct Eq.\ref{FDT_Elastico} to  yield Eq.\ref{rettaelastica}. 
{  Before to start, it is worth noting that discreteness is apparent in the elastic response of the particle position  if the elasticity is weak and $\langle u^2\rangle$ is large. Instead, if the rigidity increases and $\langle u^2\rangle$ tends to vanish, say $\langle u^2\rangle \ll 0.1$, Eq.\ref{rettaelastica} reduces to Eq.\ref{FDT_Elastico}. From this respect, we are in harmony with  the microscopic single particle barrier hopping theory of glassy dynamics which in the same limit, dubbed "ultralocal", also derives Eq. \ref{FDT_Elastico} \cite{SchweizerUltraLocJCP07,note3}.}

\begin{figure}[t]
\begin{center}
\includegraphics[width=4cm]{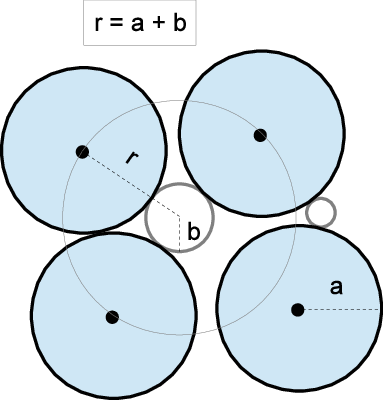}
\end{center}
\caption{ Mimicking the spontaneous local expansion of a liquid by particle insertion. A particle of radius $b$ (white) inserted in a liquid of particles  with radius $a$ (blue) creates a "cavity" of radius $r=a+b$, excluding the centers of the blue particles. The small white particle on the right has radius $b^\dag = 0.17 a$. According to the elastic cavity model, the expansion following its insertion is enough for the relaxation of the liquid.  That expansion is too small to be dealt with by the elastic continuum limit, which requires $b \gg a$. }
\label{fig0}
\end{figure}

\subsection{Elastic cavity model.}
\label{elasticcavity}
The inverse of the mean square rattling amplitude of the monomer in the cage $1/ \langle u^2\rangle$ is a measure of the average activation free-energy barrier $\Delta F^\dag$  for structural relaxation  \cite{DyreRevModPhys06,HallWoly87}:
\begin{equation}
\frac{1}{\langle u^2\rangle} = \frac{2}{3 \, k_B T \, {r_0^2 }} \, \Delta F^\dag
\label{HW}
\end{equation}
where $r_0$ is the average distance to get to the transition state (for the present polymer model $3 r^2_0 /2 = {\beta} \ln 10 \simeq 0.061$, where the  $\beta$ parameter is taken from ref. \cite{wolyNatPhys}). We interpret the barrier $\Delta F^\dag$ as the reversible work $W(0 \to R^\dag)$ to bring about a local expansion and create a cavity with radius $R^\dag$ \cite{DyreRevModPhys06}:
\begin{equation}
\Delta F^\dag = W(0 \to R^\dag)
\label{HW2}
\end{equation}
It must be noted that,
while the expansion occurs in any EC site with {\it equal} probability, the cavity nucleates only {\it outside} the particles in a {\it discrete} ensemble. This results in an entropic barrier which adds to the elastic one. 

To estimate the effect, we consider the simplest discrete ensemble of particles, i.e. a liquid of hard spheres. In this case the work $W(0 \to r)$ has been evaluated by Reiss {\it et al.} in the framework of the so called scaled particle theory (SPT) \cite{Lebowitz59,Lebowitz60}. SPT defines a cavity of radius $r$ as an empty domain being able to exclude the centers of other particles from a region of radius $r$, see Fig.\ref{fig0}. The insertion of a sphere of radius $b$ in a liquid of molecules with radius $a$ is equivalent to the creation of a cavity of radius  $r = a+b$   \cite{Lebowitz60}. 
$W(0 \to r)$ is conveniently written as:
\begin{equation}
W(0 \to r) =  W(0 \to a)  +  W( a \to r) \hspace{8mm} r \ge a
\label{HW3}
\end{equation}
where $W(x \to y)$ is the work to expand the cavity radius from $x$ to $y$. The term $W(0 \to a)$ is written by SPT as \cite{Lebowitz59,Lebowitz60}:
\begin{equation}
 W^{SPT}(0 \to a) = - k_B T \ln [ 1 - \frac{4}{3} \pi \rho a^3]   
\label{HW4}
\end{equation}
$\rho$ is the number density. The argument of the logarithm expresses the probability that the center of the cavity is located in the available space between the particles. 
{ For $r \ge a$, SPT writes the term $W( a \to r)$ of Eq.\ref{HW3} as (see Eq. 1.8 of ref. \cite{Lebowitz60}):
\begin{equation}
W^{SPT}( a \to r) =  k_1 (r-a) + k_2 (r-a)^2 + k_3 (r-a)^3
\label{HW5}
\end{equation}
$k_1$, $k_2$  are constants being set by requiring that the first and second derivatives of $W(0 \to r)$ are continuous at $r=a$ (note that $W(0 \to r) = - k_B T \ln [ 1 - 4/3 \pi \rho r^3]$ for $r\le a$, see Eq. 1.4 of ref. \cite{Lebowitz60}), whereas $k_3$ is related to the external hydrostatic pressure.}
By neglecting the volume work against the external hydrostatic pressure {($k_3=0$), a safe assumption for liquids under normal conditions}, SPT expresses the limit form of $W^{SPT}( a \to r)$ for large cavities in terms of the surface work as \cite{Lebowitz59,Lebowitz60}:
\begin{equation}
W^{SPT}( a \to r) =  4\pi r^2 \gamma \left ( 1 - \frac{2 \delta}{r} \right ), \hspace{5mm} r \gg a
 \label{SPT}
\end{equation}
$\gamma$ is a planar surface free-energy, i.e.  the interfacial tension between the bulk liquid and the cavity in the limit of infinite radius. The factor $( 1 - 2 \delta/ r )$ corrects the surface free-energy for the finite curvature of the interface, where $\delta$ is the Tolman length which is of the order of the thickness of the layer near the interface.

We propose to write the term $W( a \to r)$ in Eq. \ref{HW3} as:
\begin{equation}
W( a \to r) =  8 \pi G_p  a (r-a)^2, \hspace{5mm} r \ge a
\label{elastic3}
\end{equation}
The term on the right hand side accounts for the elastic energy if the expansion is performed preserving local mechanical equilibrium \cite{Frenkel}. Mechanical equilibration is {\it completed} in our polymer model in times shorter than $t^\star$, the time scale  where the modulus $G_p$ and the position fluctuations $\langle u^2\rangle$ are evaluated, see Sec.\ref{SimMD}. {Eq.\ref{elastic3} has the form of Eq.\ref{HW5} with $k_1=k_3=0$ and}  for large cavity, $r \gg a$,  recovers Eq. \ref{SPT} with  $\gamma =  2 G_p a$ and $\delta = a$.

By plugging Eq.\ref{HW4} and Eq.\ref{elastic3} into Eq. \ref{HW3} and resorting to Eqs.\ref{HW} and \ref{HW2} one recovers Eq.\ref{rettaelastica} with:
\begin{eqnarray}
\frac{1}{\zeta^2} &\equiv&  \frac{2}{3 r_0^2} \ln \left [ \frac{1}{1 - \frac{4}{3} \pi \rho a^3} \right ]   \label{rettaelastica1}\\
\ell &\equiv& \frac{2}{3 r_0^2} \; 8 \pi  a (R^\dag-a)^2
\label{rettaelastica2}
\end{eqnarray}

The  EC limit  is reached by setting $ a \ll b^\dag, \rho^{-1/3}$ ($b^\dag \equiv R^\dag-a$). In this case one approximates $\Delta F^\dag = W(0 \to R^\dag) \simeq W( a \to R^\dag) \propto G_p$ and  Eq.\ref{FDT_Elastico} is recovered. 

The cavity model fits with the MD results far from the EC limit. To check this, we notice that the best-fit values of Eq.\ref{rettaelastica} to the MD results (Fig.\ref{fig1}top) correspond to $R^\dag \simeq 1.17 \, a$ and $a \simeq 0.4$  by taking $\rho = 1.05$ as typical density (the $a$ radius compares well with the effective monomer radius $\sim 0.48$ estimated as in Ref.\cite{StarrEtAl02}).  Then, the local expansion involved in the relaxation, $b^\dag = (R^\dag-a) \simeq 0.17a $,  is too small to be dealt with by the continuum picture ($a \ll  b, \rho^{-1/3}$), see Fig.\ref{fig0}. 

{ Interestingly, the characteristic length $\ell$, Eq.\ref{rettaelastica2},  has been derived as \cite{SchweizerElastic09}:
\begin{equation}
\ell ' = \frac{5  }{3 \sqrt{\pi} \rho \sigma^2} 
\label{FDT_ElasticoKS}
\end{equation}
with $\rho = 1.05$ and $\sigma \sim 2 a \sim 1$ one finds $\ell' \simeq 0.85$, to be compared with our best-fit value $\ell = 0.77$.}

\begin{figure}[t]
\begin{center}
\includegraphics[width=8cm]{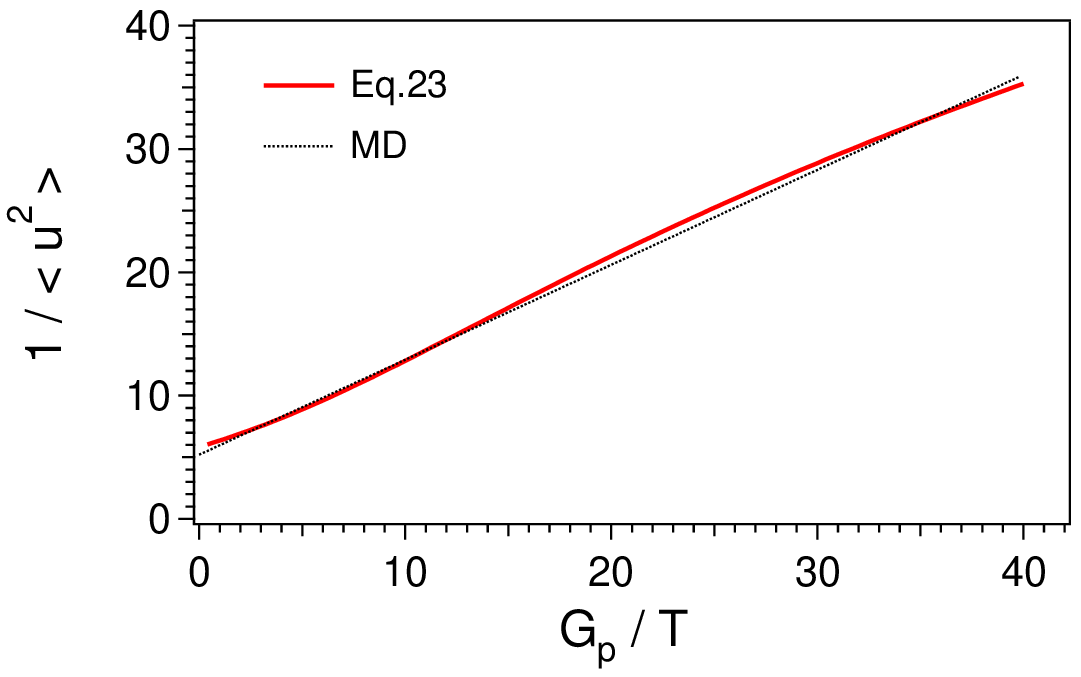}
\end{center}
\caption{Comparison of  the best-fit line of the MD results in Fig.\ref{fig1}(top) and Eq.\ref{Ziman2} with $\epsilon = 0.13$, $\overline{R} = 1.018$, $a=0.5$. The $C$ parameter is adjusted to ensure coincidence of the two curves at $G_p/T = 10$. }
\label{figX}
\end{figure}

{  The cavity model interprets, via Eq.\ref{rettaelastica1},  the characteristic length scale $\zeta$ as due to packing effects, so that the mean square rattling amplitude in the cage $\langle u^2\rangle$ is seen to be affected by both the local free volume and the cage softness. This remark suggests a simplified version of the cavity model. 
Suppose that the particle of radius $a$ is located in a cavity of radius $R \gtrsim  2a $ ($R$ must not be confused with the radius $r$ of the cavity involved in the local expansion, see Fig.\ref{fig00}). The centre of the atom traces out a free volume $v_f \propto (R-2a)^3$ \cite{ZimanBook,LennardJonesDevonShire37,StarrEtAl02}. One expects that the rattling amplitude in the cage is proportional to $v_f^{2/3}$ \cite{HallWoly87,StarrEtAl02,DouglasStarrPNAS2015}: 
\begin{equation}
\langle u^2\rangle_R =  C (R-2a)^2  
\label{Ziman}
\end{equation}
where $C$ is a constant and the subscript reminds that the average has to be intended at fixed $R$.  We take the cavity radius as a quantity elastically fluctuating around the average size $\overline{R}$  on much slower time scale than the time needed by the trapped particle to rattle in the allowed free volume.  Then, $\langle u^2\rangle$ is a weighted average over the distribution of the cavity size:
\begin{equation}
\langle u^2\rangle =  \frac{1}{{\cal N}} \int_{2a}^{2a (1 + \epsilon)} C (R-2a)^2 e^{- \beta W_c} R^2 d R
\label{Ziman2}
\end{equation}
where ${\cal N} = \int_{2a}^{2a (1 + \epsilon)} \exp[- \beta W_c ] R^2 d R $ and $W_c$ is the elastic energy of a cavity with radius $R$, $W_c = 8 \pi G_p  \overline{R}  (R- \overline{R})^2 $\cite{Frenkel}. Eq.\ref{Ziman2} takes into account that the fluctuations of the cavity size occur between the particle diameter $2a$ and a quantity slightly larger, $2a (1 + \epsilon)$, due to high packing. 
Fig.\ref{figX} compares the simplified model  with the best-fit line of the MD results. The best-fit values of the model parameter comply with some expected constraints, namely $\epsilon \sim  b^\dag /a = 0.17$ and the inequalities $2a \le \overline{R} \le 2a (1 + \epsilon)$.

The two models that we discussed are rather different from each other but they share the common assumption that the mean square rattling amplitude $\langle u^2\rangle$ is affected by both the available free volume and the softness of the surroundings. The fact that both models consistently support Eq.\ref{rettaelastica} suggests the robustness of this hypothesis and strengthen their interpretation of the characteristic length scale $\ell$ appearing in Eq.\ref{rettaelastica} as a free-volume effect not accounted for by the EC description.  
}

\begin{figure}[t]
\begin{center}
\includegraphics[width=8.5cm]{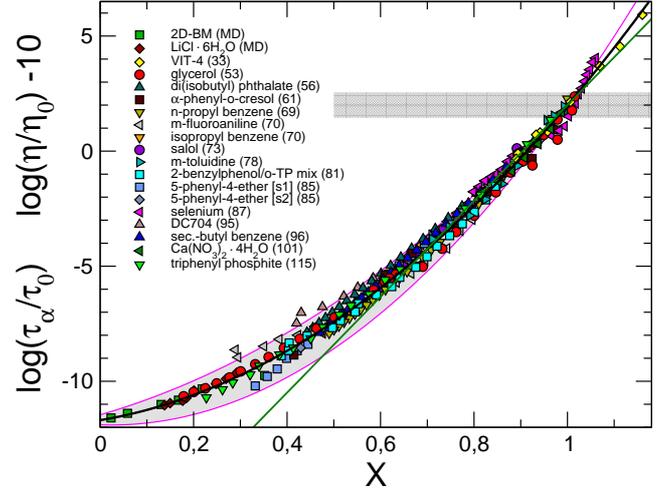}
\end{center}
\caption{ Scaling of the structural relaxation time and viscosity in terms of the reduced variable $X = G_p  T_g/(G_{pg} T)$. The numbers in parentheses denote the fragility of the glassformers. Simulation data concerning  atomic (2D-BM)  and  ionic (LiCl-6H${}_2$O) glassformers are also included. All data sources are listed in SI. 
Note that the set of glassformers under consideration widely {differs} from the one assessing Eq.\ref{parabolaMD}\cite{PuosiSpecialIssueJCP13}. The black solid line is the master curve given by Eq.\ref{universalelastic} with errors bounded by the magenta lines. 
The experimental data concerning the relaxation time and the viscosity are adjusted to fit with Eq.\ref{universalelastic} by the vertical shifts $ \log \tau_0$ and $\log \eta_0$, respectively, which are both less than $0.4$ in magnitude. No other adjustment is done.  The MD data are shifted by $\log \tau_0=11.5$, to be interpreted as the (fixed) conversion factor between MD and SI time units \cite{wolyNatPhys}. 
All the vertical shifts are listed in SI. 
The elastic moduli measured at frequency $\omega$ are considered only in the temperature region where $\omega  > max \{ 0.1/\tau_\alpha, 2\pi \cdot 1 kHz\}$.
The green straight line is the best-fit with Eq.\ref{ElasticModel} written in terms of $X$ and having adjusted {\it two} parameters: the slope $V^\star \, T_g/G_{pg}$ and  $ \tau_{\alpha 0}$. 
}
\label{fig2}
\end{figure}

\subsection{Experimental evidence of the elastic scaling.}
We recast the MD master curve, Eq.\ref{masterG}, in a scaled form by considering   
the reduced variable $X$ defined in Eq.\ref{reducedX} ($T_g$ is defined by the familiar relations $\tau_\alpha(T_g) = 100$ s or $\eta(T_g) = 10^{12}$ Pa$\cdot$s corresponding to $\tau_\alpha = 10^{13.5}$ in MD units \cite{wolyNatPhys}). We obtain:
\begin{equation}
\log \frac{\tau_\alpha}{\tau_0} = \widetilde{\Upsilon}_0 + \widetilde{\Upsilon}_1 X + \widetilde{\Upsilon}_2 X^2
\label{universalelastic}
\end{equation}
where $\widetilde{\Upsilon}_1$ and $\widetilde{\Upsilon}_2$ are deemed to be universal constants and $\tau_0=1$.  { $\widetilde{\Upsilon}_1$ and $\widetilde{\Upsilon}_2$ are derived as follows}. From the best-fit of Eq.\ref{rettaelastica} to the MD data (Fig\ref{fig1} top), and reminding that $\langle u^2 (T_g) \rangle = 0.0166$ in MD units \cite{wolyNatPhys},  one finds $G_{pg}/T_g = 71.3$ in MD units. 
Then, one finds with $\tau_0=1$, $\widetilde{\Upsilon}_0  = -11.70(1), \widetilde{\Upsilon}_1 = \Upsilon_1 G_{p g}/T_g = 3.4(2), \widetilde{\Upsilon}_2 = \Upsilon_2 (G_{p g}/T_g)^2 =  10.3(8)$.  $ \Upsilon_1$ and $ \Upsilon_2$ are taken from Fig.\ref{fig1}.  The $\widetilde{\Upsilon}_0$ parameter is set so as to get  $\log \tau_\alpha  = 2$ at $T_g$.

Fig.\ref{fig2} shows the elastic scaling and the comparison with the {\it fragility-independent} master curve, Eq.\ref{universalelastic}, for several glassformers spanning a wide range of fragilities ($33 \le m \le 115$, sources in SI). { Note the most fragile glassformer, decaline, has $m \sim 145-147$  
\cite{DecalineFragility}}.
The effectiveness of the elastic scaling reveals that, as it happens in MD simulations, the temperature dependence of $\log \tau$ and $\log \eta$ is highly correlated with the one of the elastic modulus.

Fig.\ref{fig2} compares the results also with the prediction of the conventional elastic models \cite{DyreRevModPhys06,Nemilov06}, Eq.\ref{ElasticModel}. It is seen that Eq.\ref{ElasticModel}  fits the scaled data around the glass transition ($-6 \le \log \tau_\alpha \le 6$) but departs when relaxation is faster in spite of two adjustable parameters (the slope $V^\star \, T_g/G_{pg}$ and  $ \tau_{\alpha 0}$). 

\begin{figure}[t]
\begin{center}
\includegraphics[width=8.5cm]{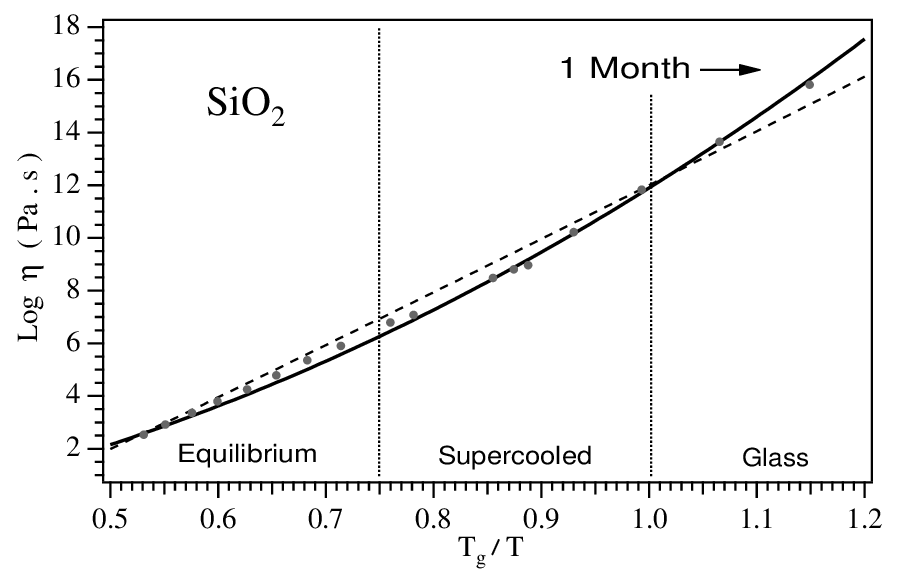}
\end{center}
\caption{ Predicted viscosity of SiO${}_2$ ($T_g = 1463 K$) in the temperature range 1273-2773 K.  The vertical dashed lines mark the melting temperature $\simeq$ 1950 K and $T_g$, separating the whole range in the equilibrium, supercooled and glass regions. The grey points are the most reliable viscosity data from 1273 to  2773 K \cite{DoremusSiO2}.
The black solid curve is the combination of the elastic parameter $X(T)$ from ref. \cite{Rouxel11} and Eq.\ref{universalelastic}  vertically shifted by $+10$ to cross the point $\log \eta= 12$ at $T_g$. No other adjustment is done.  The dashed line is the best-fit of the experimental data provided by Eq.\ref{ElasticModel} { having adjusted both $\tau_{\alpha 0}$ (to cross the point $\log \eta(T_g) = 12$) and $V^\star$}. At $\log \eta  \simeq 16.4$ the structural relaxation time is about 1 month.
}
\label{fig4}
\end{figure}

\subsection{Prediction of the silica viscosity.} 

To assess the predictability of Eq.\ref{universalelastic}, we evaluate the viscosity of the strong glass former SiO${}_2$ (fragility index $m=20$) on the basis of the sole elasticity data \cite{Rouxel11}. Then, we compare the result with the recommended viscosity values measured over a wide range of temperatures from the glassy state up to the equilibrium  where the non-Arrhenius dependence becomes apparent \cite{DoremusSiO2}. The test is severe in that, up to now, both MD simulations and experimental data concern liquids with high and intermediate fragility, whereas silica is extremely strong. 
The results are in Fig.\ref{fig4}.  Apart from adding a vertical shift to Eq.\ref{universalelastic} by $+10$ to ensure $\log \eta(T_g) = 12$, no adjustment is done. We find excellent agreement over about fifteen decades  from below $T_g$ up to liquid states. 
It is known that the conventional elastic models work well for strong glassformers \cite{Rouxel11}. However, in spite of the adjustable parameter $V^\star$ ($\tau_{\alpha 0}$ is set by the constraint $\log \eta(T_g)$ = 12), Eq.\ref{ElasticModel} is unable to account for the non-Arrhenius dependence and exhibits larger deviations than our prediction.

\section{Conclusions}

Simulation results compared with relaxation data  covering eighteen decades in glassformers with widely different fragilities ($33 \le m \le 115$) show that relaxation and linear elasticity scale to a fragility-independent master curve. The scaling allows to derive the viscosity of supercooled silica ($m=20$) over about fifteen decades with no adjustable parameters. The elastic scaling is related to the previously reported scaling between the fast mobility and the structural relaxation by a cavity model interpreting the rattling motion of a particle in the cage of the first neighbours as affected by both the available free volume and the softness of the cage. This picture appears to be robust.

{ The paper strongly suggests that the kinetic fragility just reflects the degree of elastic softening, i.e. the decrease of the elastic modulus with increasing temperature. This provides a connection between the thermodynamics, via $G_p$, and the kinetics.}

\begin{acknowledgement}
S.Capaccioli and A. Ottochian are warmly thanked for several discussions. A generous grant of computing time from IT Center, University of 
Pisa and ${}^\circledR$ Dell Italia is gratefully acknowledged.
\end{acknowledgement}

%
\bibliographystyle{epj}

\begin{thebibliography}{67}
\bibitem{note3}
The ultralocal limit is expected to hold if the localisation length of the
  theory $r_L \ll \sqrt{3} / k_c \sim (1.4 \sigma ) \sqrt{3} / 2 \pi $
  \cite{SchweizerUltraLocJCP07,SchweizerElastic1JCP14}. Since $r^2_L = \langle
  u^2\rangle$, this implies $\langle u^2\rangle \ll 0.15$.
\end{thebibliography}

\end{document}